\documentstyle[prd,aps,graphicx]{revtex}
\begin{document}
\draft
\title{On convergence of the $\chi$PT HFF expansion
for one loop contribution to
meson production in NN collisions
}
\author{E.Gedalin\thanks{gedal@bgumail.bgu.ac.il},
 A.Moalem\thanks{moalem@bgumail.bgu.ac.il}
 and L.Razdolskaya\thanks{ljuba@bgumail.bgu.ac.il}}

\address
{Department of Physics, Ben Gurion University, 84105, Beer Sheva,
Israel}
\maketitle

\begin{abstract}
We consider the application of heavy fermion formalism based
chiral perturbation theory to meson production in
nucleon-nucleon collisions. It is shown that for one loop 
contributions the heavy fermion formalism expansion
corrections for the nucleon propagator produce  infinite series of
correction terms which are of the same momentum power order. This
destroys the one-to-one correspondence between the perturbative
and small momentum expansion and thus negates the application of
any finite order heavy fermion formalism chiral perturbation theory 
to the $NN\rightarrow NN \pi$ reactions.

\end{abstract}


\section{Introduction}
In recent years, intensive theoretical efforts have been
devoted to investigating  how nuclear and hadron interactions
can be understood within Chiral Perturbation Theory ($\chi$PT),
an approach which is generally believed to be an effective
theory of Quantum Chromodynamics (QCD) at low energies.
Hopefully, such an approach will provide a clue towards
understanding nuclear dynamics within the context of QCD,
the accepted fundamental theory of strong interactions.

In accordance with the Goldstone theorem, the low energy strong
interactions of colorless particles are dominated by colorless
pseudoscalar massless boson octet. Since all interactions of the
Goldstone bosons vanish at zero energy, the amplitude of processes
(between hadrons) can be expanded in powers of external momenta
and quark masses. This expansion amounts to a derivative expansion
of the effective Lagrangian which is based on the nonlinear
realization of chiral symmetry, so that perturbative expansions of
effective field theories reproduce the low momentum expansion of
the QCD Green functions. A drawback of this fully relativistic 
approach is that no one-to-one correspondence between loop 
contributions and small momentum expansion can be established 
for processes with nucleons. 
Such a correspondence is believed to be restored in extremely
non-relativistic approximation of the heavy fermion formalism
(HFF) based $\chi$PT\cite{park93,bernard95}. In particular, Cohen
et al. \cite{cohen96} proposed a modified power counting for meson 
production in NN collisions, and a 
scheme as such was used by several groups\cite{park96}-\cite{gedalin99} to
calculate pion production rate in nucleon-nucleon collisions.
Namely, Park et al.\cite{park96}, Cohen et al.\cite{cohen96} and
Sato et al.\cite{sato97} have considered the $pp \rightarrow pp
\pi^0$ process taking into account contributions from lowest order
tree graphs. The leading order impulse and rescattering terms are
found to have opposite signs, and hence leading to a cross section
substantially smaller than experiment\cite{park96,cohen96,sato97}.
Much of the virtue of these
calculations resides on how rapid the HFF expansion converges.
Detailed $\chi$PT calculations which account for all contributions
from tree and one loop diagrams to chiral order D=2, show that
within the framework of the HFF, one loop contributions are
sizably bigger than the lowest-order impulse and rescattering
terms\cite{gedalin98,rocha99}, indicating that the HFF power
series expansion converges rather slowly and therefore may not be
suitable to calculate pion production rate in NN
collisions\cite{gedalin98}. More recently Bernard et
al.\cite{bernard98} and Gedalin et al.\cite{gedalin981} have shown
that the HFF power series expansion of the nucleon propagator is
on the border of its convergence circle, and concluded that a
finite order HFF can not possibly predict nucleon pole terms
correctly and hence  can not explain meson production in NN
collisions.

In this regard, it is to be noted that meson-exchange
models as well as a fully relativistic $\chi$PT predict
equal signs for the impulse and rescattering terms, achieving quite
impressive descriptions of data near threshold
\cite{hernandez95,hanhart95,gedalin96,gedalin98,hanhart98,tamura98,gedalin981}.
Particularly, in covariant exchange models\cite{hernandez95,gedalin98}
the amplitudes from the impulse and rescattering terms
interfere constructively. It has been argued\cite{hanhart98} 
that this sign difference between predictions of meson exchange 
models and HFF $\chi$PT is a genuine feature. Tamura et
al. \cite{tamura98} have shown that the shape of energy spectra for 
the $d(p,(pp)_s)\pi^0 n$  and $d(p,(pp)_s)\pi^- p$ reactions, 
can be explained only if the interference between these terms is 
constructive. Next, Gedalin et al.\cite{gedalin981} have shown 
that in a relativistic $\chi$PT the sign and relative importance of
various contributions are different from those found using  HFF;
the rescattering term is found to be as large and having an 
equal sign as the impulse term. It was also shown that these 
differences are mostly an outcome of the fact
that the HFF Lagrangian provides a reasonable nucleon kinetic term
only for energies close to zero and therefore, the validity of
the HFF Lagrangian becomes limited to rather small nucleon momenta.

It is the purpose of the present note to show that within the 
framework of  HFF $\chi$PT and for sufficiently large 
momentum transfer processes, the one-to-one correspondence between 
the perturbative expansion and small momentum expansion is
badly destroyed at tree level as well as for one loop contributions.
Namely, the HFF expansion corrections for the nucleon propagator 
form an infinite series of terms all of which having
the same  momentum power. In Sec. II we examine the  influence of
HFF nucleon propagator corrections to tree and one loop diagrams.
In Sec. III we apply power counting scheme for further examination
of the validity of the HFF to pion production in NN collisions. We
conclude in Sec. IV.

\section{Tree and Loop Contributions}
To be specific we consider pion production via the $NN \to NN \pi$
reaction. In the two flavor sector of HFF $\chi$PT, the important 
contribution to the meson production amplitude comes from 
the (irreducible) diagram 1a, where a virtual pion created on one 
of the incoming nucleons is converted into a final pion on the 
second nucleon via a $\pi N \rightarrow \pi N$ conversion process.
 The characteristic four momentum transferred at threshold is 
$q \approx (-m/2,\sqrt {M m})$, with $M$ and $m$ being the masses 
of the nucleon and meson produced. Here to be emphasized that
the corresponding $\pi N \rightarrow \pi N$  amplitude is
half off mass shell and involves large momentum transfer.
This as will be demonstrated, constitutes a major difficulty
implementing HFF $\chi$PT to the production process.

As already mentioned above, in the transition from
a relativistic to non-relativistic HFF Lagrangian one reduces
the nucleon kinetic term. This  affects the nucleon propagator
strongly and in turn tree level as well as one loop contributions.
We demonstrate that by considering u and s nucleon pole terms
(diagrams 2a and 2b) and one loop (diagram 2c) contributions
to the $\pi N \rightarrow \pi N$ conversion amplitude. All of
these involve the nucleon propagator,
\begin{equation}
S_N(p) = i \frac {p\! \! / + M}{p^2 - M^2}~,
\label{propag}
\end{equation}
which we may separate into a positive and negative energy parts.
To this aim we write the numerator in Eqn. \ref{propag} as
\begin{eqnarray}
& & p\! \! / + M = M (1 + v\! \! \! /) + (p\! \! / - Mv\! \! \! /)
= 2M P_+ + l\! \! /(P_+ + P_-)~,
\end{eqnarray}
where $v^{\mu}$ denotes the four velocity of the nucleon with $v^2 = 1$,
$l^{\mu} = p^{\mu} -M v^{\mu}~$ is its residual momentum,
and $P_{\pm} = (1/2)(1 \pm v\! \! \! /)$ are operators which
project the nucleon Dirac field into large and small components.
Following the usual relativistic to non-relativistic reduction
procedure(see for example\cite{park93}), we define
$p^{\mu}_{\bot} = p^\mu -(pv)v^\mu$ to be the transversal component
of the nucleon momentum and write the nucleon propagator in the form,
\begin{equation}
S_N(l) = i \left[2M P_+ + l\! \! \! / (P_+ +P_-)\right]
\frac {1}{2 (M + T(p_{\bot}))}\left[\frac {1}{vl - T(p_{\bot})} -
\frac {1} {2M +vl + T(p_{\bot})}\right]~,
\label{propagator}
\end{equation}
with $T(p_{\bot}) = \sqrt {M^2 + p_{\bot}^2} - M$ .
( The expression, Eqn. \ref{propagator},
is just a solution of the HFF equations for the 
nucleon propagator quoted by Park, Min and Rho \cite{park93}).
By making use of this expression the contributions from s and u
channels (e.g., graphs 2a and 2b) separate into negative
and positive energy parts. The former part behave like 
a regular contact term. For example, the non-relativistic limit of
the negative energy part of the nucleon propagator at low kinetic 
energy reduces to $S_N \approx - i P_+ /{2M}$, and  the respective
negative energy contributions from graphs 2a and 2b sum up to
be,
\begin{equation}
T_Z \approx \frac {g_A^2}{4MF^2} (vq)(vk)~.
\label{contact}
\end{equation}
This indeed has the form of a contact term. In passing by, we note
that by separating the nucleon propagator into negative and positive
energy parts the s and u channels (graphs 2a, 2b) split into direct and
Z-graph contributions\cite{alfaro}. In the non-relativistic limit
the Z-graph contribution appears in exactly the form of Eqn.
\ref{contact}, as a local rescattering term, e.g. the negative energy 
part of the nucleon pole terms "converts" into a sea-gull contact term.

A serious difficulty in the reduced non-relativistic limit concerns
with the non-local positive energy part of nucleon pole terms. It is not
always possible to calculate 
the positive energy part within 
the frame of HFF. To see this consider the expansion of $S_N$, 
Eqn. \ref{propagator} as "low momentum" power series. The series 
expansions of  the factors $1/(M + T(p_{\bot})$ 
and $1/(2M + vl + T(p_{\bot}))$ converge up to 
sufficiently high energies. However, the series
\begin{eqnarray}
\frac {1}{vl - T(p_{\bot}) + i \epsilon} = \frac {1}{vl + i\epsilon}
\sum \left[ \frac {T(p_{\bot})}{vl + i\epsilon}\right]^n~,
\label{serprop}
\end{eqnarray}
converges only for  $T(p_{\bot}) / vl < 1$.
In the instance of $\pi N \rightarrow \pi N$ scattering and in the limit
$T(p_{\bot}) \ll vl$ this series converges rather well.
 However for a production process $NN \rightarrow NN \pi$,
the virtual nucleon in the graph 2b has  a residual momentum
$l' = (-m/2, {\bf l'}) ;\quad\mbox{with}\quad{\bf l'} \cdot {\bf l'} = Mm$,
so that $T(p_{\bot}) / vl' \approx -1$. Thus the corrections
corresponding to each of the terms in Eqn. \ref{serprop} 
are all of the same magnitude, i.e.,the power series of the nucleon
propagator in the graph 2b is on the border of its convergence
circle and can not be approximated by any finite sum. Consequently, the 
HFF can not possibly predict the u channel impulse term correctly.

This same conclusion holds for one loop diagrams also.
As an example the contribution from the loop diagram of
Fig. 2c to the off mass shell $\pi N$ elastic scattering amplitude
has the form,
   \begin{eqnarray}
 &  & T
_{loop} =  \frac{2g^2_A}{3F^4}
\left[3Q^2 - q^2 - k^2 + \frac{m^2}{2}\right]
N^\dagger (p_2) S^\nu S^\mu N(p_1)
   \nonumber \\
 &  & \int Dq' q'_\mu (q'_\nu + Q_\nu) \tilde{S}_N(p_2-q')
\left[(q'^2 -m^2)[(q' + Q)^2 - m^2]\right]^{-1}~,
   \label{amp}
   \end{eqnarray}
where $\tilde{S}_N(l)$ stands for the positive energy part of
the nucleon propagator, i.e.,
  \begin{equation}
\tilde{S}_N(l) = i\frac{1}{vl + i\epsilon}\sum_{n}
\left(\frac{(vl)^2 -l^2}{2M(vl + i\epsilon)}\right)^n
+ \ldots~.
  \label{propn}
  \end{equation}
In our notations
   \begin{equation}
 Dq' = \frac{\lambda^{4-d}}{(2\pi)^d}dq'~,
   \label{€}
   \end{equation}€
with $\lambda$ being the scale of dimensional regularization,
$k$ and $q$ are four momenta of the incoming and outgoing pions
(see Fig. 2c), $Q \equiv (Q^0,\vec Q)$ is the transferred momentum
$ Q^2 = (p_2-p_4)^2 = (v Q)^2 - {\vec Q}^2 $,
$S^\mu$ is the nucleon spin-operator and  F and $g_A$ are
the pion decay and axial-vector coupling constants, respectively.

To simplify calculations, we take $\vec{p}_4=0,~~~ Q^2<0,
 ~~~v Q =0 $, values corresponding to the reaction $NN
 \rightarrow NN\pi$ at threshold. After straightforward though long and
tedious calculations, the $n$-th correction term of the amplitude, Eqn.
\ref{amp} is,
 \begin{eqnarray}
  &  & T_{loop}^{(n)} =  \frac{g^2_Am}{6F^4}\left[3Q^2 - q^2 - k^2 +
  \frac{m^2}{2}\right]N^\dagger (p_2)N(p_1)
 \nonumber \\
  &  &\frac{i}{16\pi^2}\left[A_1(Q^2/m^2) +
 \frac{-Q^2}{m^2}\left(A_2(Q^2/m^2) + A_3(Q^2/m^2)\right)\right]~,
 \label{ampcorn}
 \end{eqnarray}€
where,
\begin{eqnarray}
     &  & A_1(X) = \left(\frac{m}{M}\right)^n
 \left(\frac{m^2}{4\pi \lambda^2}\right)^{-\epsilon}
 \frac{\Gamma(2-\epsilon)\Gamma(\frac{n+1}{2})}{\Gamma(\frac{3}{2}-\epsilon)}
    \nonumber \\
     &  & \sum_{k,r}\frac{\Gamma(\frac{5}{2}-\epsilon+n-k-r)
     \Gamma(-\frac{n+1}{2}+\epsilon+k+r)}
     {\Gamma(n-2k-r+1)\Gamma(k+3-\epsilon)\Gamma(2k+1)\Gamma(r+1)}
     \nonumber \\
     &&X^{k+r}
     F(2(k+r),\frac{n+1}{2}-\epsilon-k-r,X)~,
    \label{a1cor}€ \\
      & & \nonumber \\
     &  &  A_2(X) =-\left(\frac{m}{M}\right)^n
 \left(\frac{m^2}{4\pi \lambda^2}\right)^{-\epsilon}
 \frac{\Gamma(2-\epsilon)\Gamma(\frac{n+1}{2})}{\Gamma(\frac{3}{2}-\epsilon)}
    \nonumber \\
     &  & \sum_{k,r}\frac{\Gamma(\frac{3}{2}-\epsilon+n-k-r)
     \Gamma(-\frac{n-1}{2}+\epsilon+k+r)}
     {(\frac{n+1}{2}-\epsilon-k-r)\Gamma(n-2k-r+1)\Gamma(k+3-\epsilon)
     \Gamma(2k+1)\Gamma(r+1)}
     \nonumber \\
     &&X^{k+r}
     \frac{\partial}{\partial X}F(2(k+r),\frac{n+1}{2}-\epsilon-k-r,X)~,
 \label{a2cor}€ \\
         &  & \nonumber \\
     &  &  A_3(X) = \left(\frac{m}{M}\right)^n
 \left(\frac{m^2}{4\pi \lambda^2}\right)^{-\epsilon}
 \frac{\Gamma(2-\epsilon)\Gamma(\frac{n+1}{2})}{\Gamma(\frac{3}{2}-\epsilon)}
    \nonumber \\
     &  & \sum_{k,r}\frac{\Gamma(\frac{3}{2}-\epsilon+n-k-r)
     \Gamma(-\frac{n-1}{2}+\epsilon+k+r)(2k+1)}
     {\Gamma(n-2k-r+1)\Gamma(k+3-\epsilon)\Gamma(2k+1)\Gamma(r+1)}
     \nonumber \\
     &&X^{k+r}
     F(2(k+r),\frac{n-1}{2}-\epsilon-k-r,X)~,
     \label{a3cor}€
\end{eqnarray}
with $\epsilon = (4-d)/2$ and
\begin{equation}
    F(a,b,X) = \int_{0}^{1} dz z^a \left(1 - Xz(1-z)\right)^b~.
    \label{fabx}
\end{equation}€
 The $\Gamma$-functions in the rhs of
Eqns. \ref{a1cor}, \ref{a2cor}, \ref{a3cor}
 show that for even $n = 2n'$ the functions $A_i$ are finite for
  $\epsilon \rightarrow 0$ and that for odd $n = 2n'+1$ they have simple
   poles for such $k$ and $r$ which satisfy
 $-\frac{n\pm1}{2}+\epsilon+k+r\leq 0$. As usual divergent terms
 must be renormalized introducing corresponding counterterms.
 We do not enter here into details of the renormalization procedure,
 but note that the finite terms, divergent terms as well as 
 counterterms undergo this same small momentum expansion problem.

In the limit of small transferred momentum $-Q^2 \leq m^2$, i.e.,
$ -(Q^2/m^2) \leq 1$ and  $~F(a,b,-Q^2/m^2) \sim 1$,
the $n$-th correction term becomes of the order $\sim (m/M)^n$
in agreement with the organizing principle of HFF $\chi$PT.

The situation changes drastically at sufficiently large momentum
transfer, say at threshold of the production process where
$-Q^2 \geq Mm$. From Eqns.\ref{a1cor}-\ref{a3cor} it 
is easy to see that the terms with $r=n, ~~k=0$ can be 
dangerous as they are of order $(-Q^2/mM)^n \geq 1$. Provided 
the  corresponding functions $F(a,b,Q^2/m^2) $ do not restore
the small factor $(-m^2/Q^2)^n$, i.e. if, for example
$F(2n,-(n-1)/2-\epsilon,Q^2/m^2) \not\sim (-Q^2/m^2)^{-n}$, such terms
would violate the HFF $\chi$PT organizing principle.

Let us now estimate the functions
$F(2n,-\frac{n-1}{2}-\epsilon,Q^2/m^2)$.
For  $n=1 ~~\mbox{and}~~ n=2$ these terms are
of order of magnitude
$\sim \Gamma(\epsilon)(Q^2/Mm) ~~\mbox{and} ~~\sim
\Gamma(\epsilon)(Q^2/Mm)^{3/2}(m/M)^{1/2}$, respectively.
For $n \geq 3$ the dominant contribution to all these dangerous
terms is finite and we may set $\epsilon =0$ in order to
estimate the functions $F(2n,-\frac{n-1}{2},X)$ at large $X$.
For even $n=2n'+2$ one obtains,
 \begin{eqnarray}
    &&F(4n'+4,-1/2-n',X) =
    \int_{0}^{1} dz \frac{z^{4n'+4}} {\left(1 - Xz(1-z)\right)^{-1/2-n'}}=
    \nonumber\\
    &&2^{-4n'+3}\left(-\frac{X}{4}\right)^{-1/2-n'}\sum_{p=0}^{2n'+2}
    \sum_{r=0}^{p}\frac{(4n'+4)!}{(4n'+4-2p)!(2p)!}
    \nonumber \\
    &&\frac{p!}{(p-r)!r!}\sigma^{2r}
        \int_{0}^{1} dt (\sigma ^2-t^2)^{-1/2-n'+p-r}~,
    \label{dangfe}
\end{eqnarray}
where $\sigma^2 = 1-4/X$. In the case where the parameter 
$X>>1$, the last integral may be 
expanded  in  powers 
of $1/X$ with a dominant term being equal to $(1/2)(-X/4)^{n' +
 r - p -1/2}$. Hence the sum  in the rhs of Eqn.\ref{dangfe} can be
written in the same power form. The largest term occurs for $r=p$, i.e.,
\begin{equation}
    F(4n'+4, -1/2-n', X) \approx \frac{1}{8(2n'+1)}
    \left(-\frac{1}{X}\right)
    \label{fevn}
\end{equation}
 for all $n'=1,2,\ldots$.
Similarly,  for  odd $n = 2n' + 1, ~~n'\geq 1$ we have
\begin{eqnarray}
     &  & F(4n'+2,-n',X) =
    \int_{0}^{1} dz \frac{z^{4n'+2}} {\left(1 - Xz(1-z)\right)^{n'}}=
    \nonumber\\
    &&2^{-4n'+3}\left(-\frac{X}{4}\right)^{-n'}\sum_{p=0}^{2n'}
    \sum_{r=0}^{p}\frac{(4n'+2)!}{(4n+2-2p)!(2p)!}
    \nonumber \\
    &&\frac{p!}{(p-r)!r!}\sigma^{2r}
    \int_{0}^{1} dt (\sigma ^2-t^2)^{-n'+p-r}
    \nonumber\\
    && \approx \frac{1}{8(2n'-1)}\left(-\frac{1}{X}\right)~.
    \label{dangfo}
\end{eqnarray}

Then one may conclude that for $n \geq 3$ the main contributions to
$A_1(Q^2/m^2)$ have the same order in $-Q^2/m^2$ and write,
\begin{equation}
    A_1(Q^2/m^2) \approx
{\cal A}_1(n)\frac{m^2}{-Q^2}~.
    \label{a1f}
\end{equation}
In this last expression, the dependence on the correction order is
incorporated in the
coefficient ${\cal A}_1(n)$. For even and  odd $n$ values these are
${\cal A}_1(n)=(3/256)\Gamma((n-1)/2)\Gamma(n/2)\Gamma(n+1)$
and
${\cal A}_1(n)=(3/256)\Gamma((n-1)/2)\Gamma((n+1)/2)/(n/2-1)\Gamma(n+1)$
, respectively. we have found that at least one term
on the rhs of Eqn.\ref{a1cor} is of the order $m^2/(-Q^2)$ for all 
$n \geq 3$.
In a similar way one estimates for $n \geq 1$,
\begin{eqnarray}
     &  & A_2(Q^2/m^2) \approx {\cal A}_2
     (n)\left(\frac{m^2}{Q^2}\right)^2~,
    \label{a2f} \\
     &  & A_3(Q^2/m^2) \approx {\cal A}_3
     (n)\left(\frac{m^2}{Q^2}\right)~,
    \label{a3f}
\end{eqnarray}
where ${\cal A}_2 (n)$ and ${\cal A}_3(n)$ incorporate as above
all dependence of $A_2 ~~\mbox{and}~~ A_3$ on
$n$. By substituting
Eqns. \ref{a1f}, \ref{a2f} and \ref{a3f} into Eqn. \ref{ampcorn} one
finds that the $n$-th correction term is of order  $(Q^2/mM)^n$. At
threshold of the pion production process, $ Q^2 =
-mM$ and each correction term $T_{Loop}^{(n)}$ involves a contribution
of the same low momentum power order independently of $n$. 
This completes our proof that for
the one loop graph 2c at large momentum transfer $-Q^2>>m^2$
the one to one correspondence between small momentum expansion and
HFF corrections  is broken, thereby violating the fundamental organizing
principle of HFF $\chi$PT.

\section{Power counting considerations}

The standard Weinberg's power counting\cite{weinberg79},
where presumably the momentum transferred is of the order 
$Q^2 \approx m^2$ can not be applied directly to
meson production.
Cohen et al,\cite{cohen96} suggested a modified scheme tailored 
to the production process.
This scheme includes the following rules :\\
(i)  a $\pi NN$ vertex of zero chiral order D=0, $V^{(0)}_{\pi NN}$,
contributes a factor $ Q/F$, \\
(ii)  a pion propagator contributes a factor $(Q^2)$,\\
(iii)  a nucleon propagator $ (vQ)^{-1} \approx m^{-1}$,\\
(iv)  a $\pi NN$ vertex of chiral order D=1, $V^{(1)}_{\pi NN}$
contributes a factor $m^{3/2}/F M^{1/2}$\\
(v)  a $2\pi$NN D=1 vertex, $V^{(1)}_{\pi \pi NN}$, contributes a
factor $ k^0 Q^0/F^2 M$. \\

To account for loop contributions also, one needs three more
rules\cite{gedalin98}:\\
(vi) a loop integration contributes a factor $(Q^2/ 4\pi)^2$.\\
(vii) a  four pion vertex of zero order, $V^{(0)}_{\pi \pi \pi \pi}$,
contributes a factor $Q^2 /F^2 \sim m M/ F^2$. \\
(viii) a 3$\pi$N zero order vertex, $V^{(0)}_{\pi \pi \pi NN}$,
contributes a factor $Q/F^3 \sim \sqrt{m M} / F^3$.\\
The last two of these factors originate
from the terms $\pi^2 (\partial_\mu \pi)^2/6F^2$ and
$S^{\mu} \pi^2 \partial_\mu \pi/ 6F^3$ in the lowest-order Lagrangian,
respectively.
(see for example Eqn. 2 of Ref.\cite{gedalin98})\\

The rule (vi) deserves some comments. We follow the procedure of loop
calculations suggested by Park et al.\cite{park93} and use the dimensional
regularization scheme (DRS). Using the parameterization scheme of Park et
al.\cite{park93} and shifting the momentum variable after Wick rotation
the loop integrand assumes the form,
\begin{equation}
    K_{\{\mu\},\{\nu\}, \{\lambda\}}f(k^2)k^{d-1}dk d\Omega _D~,
\end{equation}
where the numerator $K_{\{\mu\},\{\nu\},\{\lambda\}}=\prod
k_{\mu}v_{\nu}Q_{\lambda}$ absorbs all tensor structure of the integrand.
Since $K_{\{\mu\},\{\nu\},\{\lambda\}}$ is multiplied by a product of
nucleon spin matrices $\prod S_\nu$ the only terms
which do not contain the nucleon velocity, $v_\nu$, 
may contribute. The denominator
of the loop integral is a power of ($k^2 + Q^2f_q + m^2f_m$)
in which the coefficients $f_q ~\mbox{and}~f_m$ do not depend on
$k^2,~~Q^2~~\mbox{and}~~m^2$. 
Therefore, a simple rescalling, $k^2 = Q^2z^2$,
enables to express the loop integral as a power expansion in 
$m^2/Q^2$. Since in the DRS all divergences separate
from the remaining integrals as the poles
at $\epsilon = (4 - d)/2 \rightarrow 0$, it
follows that the loop integration gives a factor $Q^4/(4\pi)^2$. In
addition, to estimate contributions corresponding to higher power of
$Q$ it is sufficient to take all internal pion momenta of 
$NN\pi$ vertices equal to $Q$. The corresponding loop integrals are
found to be finite.

Using these power counting rules the $n$-th correction term
$T_{Loop}^{(n)}$ may be estimated as follows: 
loop integration contributes a factor
$Q^4/(4\pi)^2$, the $\pi NN$ vertices a factor $Q^2/F^2$, $4\pi$-vertex
a factor $Q^2/F^2$, the pion propagators give $Q^{-4}$ and $n$-th order
correction of the nucleon propagator a factor $(1/v Q)(Q^2/2vQ)^n$.
For the one loop diagram 2c  one obtains,
\begin{equation}
T_{Loop}^{(n)} \sim \frac{Q^4}{F^2(4\pi F)^2 v Q}\left[\frac{Q^2}{Mv
Q}\right]^n \sim \frac{1}{F}~,
    \label{€}
\end{equation}€
which is of the same order of magnitude as the first (uncorrected) 
term $T_{Loop}^{(0)}$.
Again from  power counting rules we reach the conclusion that for 
large momentum transfer processes the basic organizing
principle of
HFF $\chi$PT is badly violated.

We now turn to demonstrate that for each low chiral order diagram
contributing to the $NN\rightarrow NN\pi$ amplitude there
exist infinite sequence of loop diagrams of higher chiral order, which
have the same low momentum power as the original diagram\cite{gedalin99}.
 Consider for example the diagrams shown in Fig. 3.
The  simplest nonvanishing at threshold irreducible diagram is the
impulse term, (graph 1a), corresponding to a chiral order $D=1$, 
of which the low momentum power order contribution 
is \cite{cohen96}, $\Theta _0 \sim F^{-3}(m_\pi/M)^{1/2}$.
Next we consider the irreducible chiral order $D = 3$
one loop diagram 3a which is obtained from
diagram 1a by adding two $\pi NN  \quad D = 0$ vertices, two lowest order
nucleon propagators  and one meson propagator.
From rule (i) above, a $\pi NN \quad D = 0$ 
vertex is proportional to the meson three momentum, i.e.,
\begin{equation}
    V_{\pi NN} = \frac{g_A}{F}S Q \tau,
\end{equation}
where S is the nucleon spin-operator. The two added nucleon vertices 
contribute a factor $Q^2F^{-2}$ and  the two nucleon propagators 
give $(vQ)^{-2}$ . Likewise, a meson propagator contributes 
a factor $Q^{-2}$ and the loop integral contributes a factor 
of $Q^4/(4\pi)^2$. Altogether, diagram 3b has an additional factor 
$Q^4(4\pi F vQ)^{-2}$ with respect to tree diagram 1a. 
The power factor of diagram 3a is therefore,
\begin{equation}
    \Theta_3 = \Theta_1 \frac{Q^4}{(4\pi F)^2 (vQ)^2}~.
\end{equation}
With  $4\pi F \sim M, \quad vQ \sim m$, the diagram 3a is of the same
order as the diagram 1a, i.e. $\Theta_3 = \Theta_1$, though higher in
chiral order.
 Similarly, by adding progressively, two zero order $\pi NN$
vertices, a pion propagator and two nucleon propagators, as mentioned
above, one constructs the other higher order irreducible n-loop diagrams
shown in Fig. 3.

By making use of the same power counting rules as above, the
momentum power of the n-loop diagram would be,
\begin{equation}
    \Theta_{2n+1} =
     \Theta_1\left(\frac{Q^4}{(4\pi f)^2 (vQ)^2}\right)^n =
    \Theta_1.
\end{equation}
We can thus construct
an infinite sequence of n-loop diagrams, $n=1,2,...$ of
chiral order $2n + 1$ all having the same characteristic 
momentum power as the lowest chiral order impulse diagram 1a. Quite 
obviously, such a sequence of loops, does exist
for any irreducible diagram that may contribute to the production 
process,and therefore we conclude that the small momentum power 
expansion scheme as a  basic organizing principle of
HFF $\chi$PT is badly violated in the instance of large momentum transfer.

\section{Summary and discussion}

We have considered tree and one loop diagram contributions to pion
production in NN collisions using the non-relativistic HFF. We
have demonstrated that the positive energy part of the nucleon
propagator, and in turn the corresponding positive energy
contributions to tree level impulse u channel diagrams and one
loop diagrams, are on the border of their convergence circles.
Also, the basic principle of the HFF $\chi$PT of one-to-one
correspondence between the loop and low momentum expansion is
badly destroyed. The primary production amplitude becomes the sum
over infinite sequences of diagrams. This  excludes the
possibility that a finite chiral order HFF based $\chi$PT
calculations can explain meson production in NN collisions.

\vspace{1.5 cm} 
{\bf Acknowledgments} This work was supported in
part by the Israel Ministry Of Absorption.

\newpage
\begin{figure}
\includegraphics[scale=0.6]{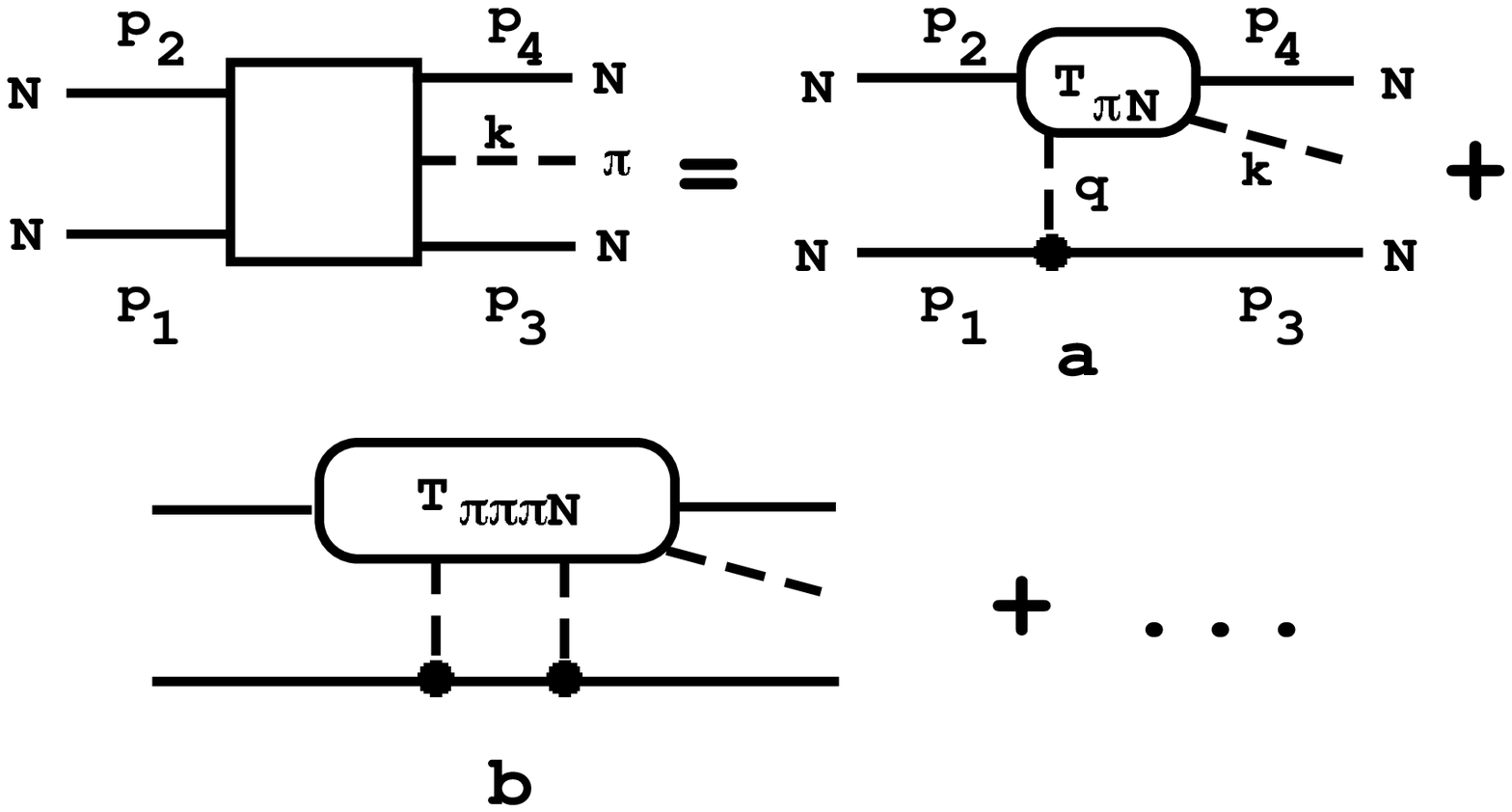}
\caption{One pion exchange ( diagram a) and two pion exchange (diagram b) 
contributing to the $NN \to NN \pi^0$
reaction. Nucleons are represented by solid lines and mesons by dashed
lines. $T_{\pi N}$ denotes the off-shell
pion-nucleon elastic scattering amplitude. $T_{\pi\pi\pi N}$ denotes the
off-shell $\pi N\rightarrow \pi\pi N$ amplitude.
}
\label{xfig1}
\end{figure}

\begin{figure}
\includegraphics[scale=0.6]{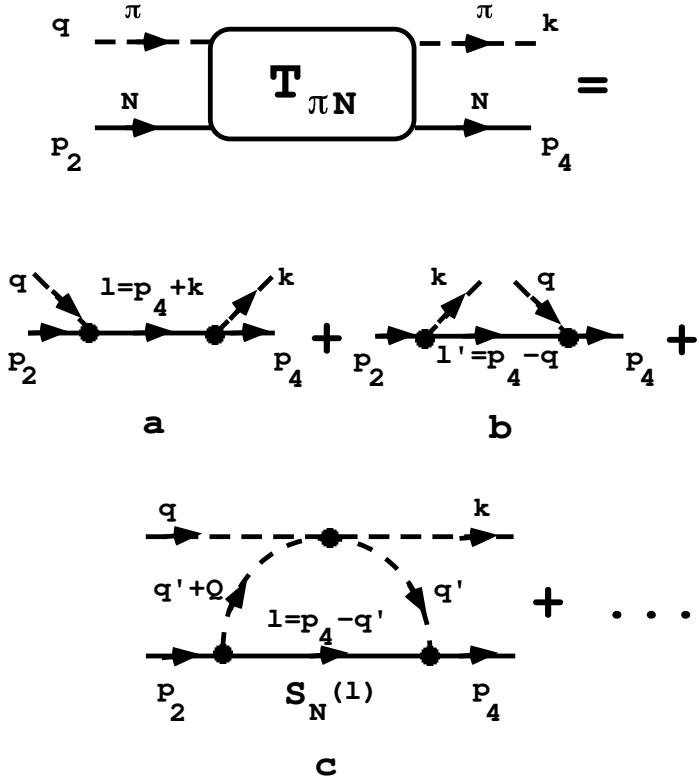}
\caption{ Pole terms (diagrams (a) and (b))and loop (diagram
c) contributions to the off mass shell 
$\pi N \rightarrow \pi N$ scattering amplitude.
}
\label{xfig2}
\end{figure}

\begin{figure}
\includegraphics[scale=0.6]{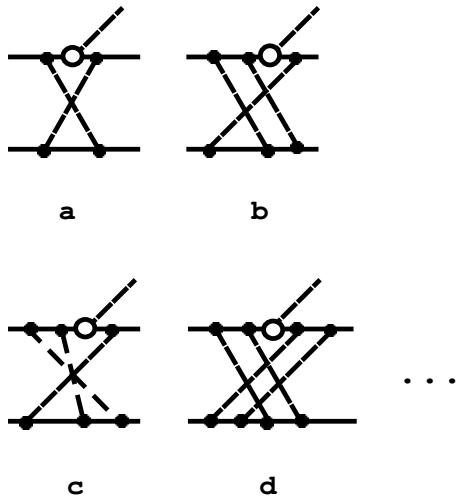}
\caption{An infinite sequence of diagrams contributing to the  pion
production process (see text). Black dots and open circle are zero
and 1 chiral order $\pi NN$ vertices, respectively.
Shown in the figure are irreducible one loop (diagram a), two-loop
(diagrams b and c) and one of the three-loop (diagram d). The ellipsis
denote all other loop diagrams of the sequence.
  }
\label{xfig3}
\end{figure}

\end{document}